\documentstyle[psfig]{mn}

\newif\ifAMStwofonts
\ifoldfss
  \ifCUPmtlplainloaded \else
    \NewTextAlphabet{textbfit} {cmbxti10} {}
    \NewTextAlphabet{textbfss} {cmssbx10} {}
    \NewMathAlphabet{mathbfit} {cmbxti10} {} % for math mode
    \NewMathAlphabet{mathbfss} {cmssbx10} {} %  "   "    "
  \fi
  \ifAMStwofonts
    \ifCUPmtlplainloaded \else
      \NewSymbolFont{upmath} {eurm10}
      \NewSymbolFont{AMSa} {msam10}
      \NewMathSymbol{\upi}     {0}{upmath}{19}
      \NewMathSymbol{\umu}     {0}{upmath}{16}
      \NewMathSymbol{\upartial}{0}{upmath}{40}
      \NewMathSymbol{\leqslant}{3}{AMSa}{36}
      \NewMathSymbol{\geqslant}{3}{AMSa}{3E}

       \let\le=\leqslant
       \let\ge=\geqslant
    \fi
  \fi
\fi % End of OFSS

\ifnfssone
  \newmathalphabet{\mathit}
  \addtoversion{normal}{\mathit}{cmr}{m}{it}
  \addtoversion{bold}{\mathit}{cmr}{bx}{it}
  \newmathalphabet{\mathbfit} % math mode version of \textbfit{..}
  \addtoversion{normal}{\mathbfit}{cmr}{bx}{it}
  \addtoversion{bold}{\mathbfit}{cmr}{bx}{it}
  \newmathalphabet{\mathbfss} % math mode version of \textbfss{..}
  \addtoversion{normal}{\mathbfss}{cmss}{bx}{n}
  \addtoversion{bold}{\mathbfss}{cmss}{bx}{n}
  \ifAMStwofonts
    \ifCUPmtlplainloaded \else
      \UseAMStwoboldmath
      \makeatletter
      \new@mathgroup\upmath@group
      \define@mathgroup\mv@normal\upmath@group{eur}{m}{n}
      \define@mathgroup\mv@bold\upmath@group{eur}{b}{n}
      \edef\UPM{\hexnumber\upmath@group}
      \new@mathgroup\amsa@group
      \define@mathgroup\mv@normal\amsa@group{msa}{m}{n}
      \define@mathgroup\mv@bold\amsa@group{msa}{m}{n}
      \edef\AMSa{\hexnumber\amsa@group}
      \makeatother
      \mathchardef\upi="0\UPM19
      \mathchardef\umu="0\UPM16
      \mathchardef\upartial="0\UPM40
      \mathchardef\leqslant="3\AMSa36
      \mathchardef\geqslant="3\AMSa3E

       \let\le=\leqslant
       \let\ge=\geqslant
    \fi
  \fi
\fi % End of NFSS release 1

\ifnfsstwo
  \DeclareMathAlphabet{\mathbfit}{OT1}{cmr}{bx}{it}
  \SetMathAlphabet\mathbfit{bold}{OT1}{cmr}{bx}{it}
  \DeclareMathAlphabet{\mathbfss}{OT1}{cmss}{bx}{n}
  \SetMathAlphabet\mathbfss{bold}{OT1}{cmss}{bx}{n}
  \ifAMStwofonts
    \ifCUPmtlplainloaded \else
      \DeclareSymbolFont{UPM}{U}{eur}{m}{n}
      \SetSymbolFont{UPM}{bold}{U}{eur}{b}{n}
      \DeclareSymbolFont{AMSa}{U}{msa}{m}{n}
      \DeclareMathSymbol{\upi}{0}{UPM}{"19}
      \DeclareMathSymbol{\umu}{0}{UPM}{"16}
      \DeclareMathSymbol{\upartial}{0}{UPM}{"40}
      \DeclareMathSymbol{\leqslant}{3}{AMSa}{"36}
      \DeclareMathSymbol{\geqslant}{3}{AMSa}{"3E}

       \let\le=\leqslant
       \let\ge=\geqslant
    \fi
  \fi
\fi % End of NFSS release 2

\ifCUPmtlplainloaded \else
  \ifAMStwofonts \else % If no AMS fonts
    \def\upi{\pi}
    \def\umu{\mu}
    \def\upartial{\partial}
  \fi
\fi

\title{Origin of the metallicity distribution of the NGC 5128 stellar halo}
\author[K. Bekki,  William E.  Harris, Gretchen L. H. Harris]
       {K. Bekki,${}^1$   William E.  Harris${}^{2,4}$, and   Gretchen L. H. Harris${}^{3,4}$\\
        ${}^1$School of Physics, University of New South Wales, Sydney 2052, NSW, Australia \\
        ${}^2$Department of Physics \&  Astronomy, McMaster University, 
Hamilton ON L8S 4M1, Canada \\
${}^3$ Department of Physics, University of Waterloo, Waterloo ON N2L 3G1, Canada\\
${}^4$ Visiting Fellow, Research School of Astronomy \& Astrophysics, Australian
National University, Weston, ACT 2611, Australia}
\date{Accepted 
      Received
      in original form 2001}

\pagerange{\pageref{firstpage}--\pageref{lastpage}}
\pubyear{1994}

\begin{document}

\maketitle

\label{firstpage}

\begin{abstract}

Recent {\it Hubble Space Telescope} photometry 
in the nearby elliptical galaxy NGC 5128 shows
that its halo field star population 
is dominated by moderately metal-rich stars, with a peak at 
[m/H] $\simeq$ $-0.4$ and with a very small fraction of metal-poor
([m/H] $<$ $-1.0$) stars.  In order to investigate the physical
processes which may have produced this metallicity distribution
function (MDF), we consider a model 
in which NGC 5128 is formed by merging of two major spiral galaxies.
%and evaluate the stellar halo MDF, particularly in terms of its
%dependence on the physical properties of the merger progenitor 
%discs (bulge-to-disc-ratio, disc's initial
%metallicity distribution, and mass fraction of the disc's stellar halo). 
We find that the halo of an elliptical formed this way 
is predominantly populated by moderately metal-rich stars 
with [m/H] $\sim$ $-0.4$ which were initially within the outer parts
of the two merging discs and were tidally stripped during the merger.
To match the NGC 5128 data, we find that the progenitor spiral discs 
must have rather steep metallicity gradients similar to the one defined
by the Milky Way open clusters, as well as sparse metal-poor haloes (5\%
or less of the disc mass).  Very few stars from the central bulges of
the spiral galaxies end up in the halo, so the results are not sensitive
to the relative sizes (bulge-to-disc ratios) or metallicities of the initial bulges.
Finally, we discuss the effects on the globular cluster system (GCS).
The emergent elliptical will end up with metal-poor halo clusters from the
original spiral haloes, but with moderately metal-rich halo stars from
the progenitor discs, thus creating a mean offset between the MDFs of
the halo stars and the GCS.
Remaining questions yet to be answered concern the total size of the
GCS population (the ``specific frequency problem'') and the observed
existence of metal-rich globular clusters in large numbers in the outer
haloes of giant ellipticals.
We also discuss possible differences
in the MDFs of stellar haloes of galaxies of different Hubble type. 

\end{abstract}

\begin{keywords}
galaxies:abundances ---
galaxies:elliptical and lenticular, cD ---
galaxies:evolution --- 
galaxies:individual (NGC 5128) ---
galaxies:stellar content ---
globular clusters:general
\end{keywords}

\section{Introduction}

Physical properties of metal-deficient stellar haloes in galaxies
are considered to provide vital clues to the understanding
of early dynamical and  chemical evolution   of galaxies. 
In particular, the detailed investigation of
structural, kinematic, and chemical properties
of such  a ``fossil record''  component (i.e., stellar halo)
of the Galaxy has revealed a possible scenario as to  how the Galaxy has developed
its dynamical structures such as bulge, thin and thick discs 
(e.g., Freeman 1987; Majewski 1993; van den Bergh 1996).
Because of  the observational difficulties in revealing three dimensional structure
and kinematics  of halo stars  (with respect to their host galaxy)
in each of  the Local Group galaxies other than the Milky Way,
the metallicity distribution function (hereafter MDF)  of halo stars in these galaxies
has served to give some constraints on 
the past star formation and chemical evolution histories of these systems 
(Mould \& Kristian 1986; Durrell, Harris, \& Pritchet 1994, 2001;
Pritchet \& van den Bergh 1998; Christian \& Heasley 1986:
Reitzel, Guhathakurta, \& Gould 1998; Grillmair et al. 1996;
Han et al. 1997; Martinez-Delagado \& Aparicio 1998).
The stellar halo of M31 appears to be dominated by a moderately high-metallicity
population with [m/H]  $\sim$ $-0.5$ (cf.~the references cited above), 
a strikingly different enrichment level than in the Milky Way stellar halo,
and a strong indicator of a different formation history (e.g.~Durrell, Harris, \& 
Pritchet 2001).

Although there have been many MDF studies of
halo stars in Local Group members,  galaxies
beyond the Local Group have only recently been investigated 
through {\it HST}/WFPC2 photometry and for only a small 
number of cases:  the giant E/S0 NGC 5128 (Soria et al. 1996;
Harris, Harris, \& Poole 1999 hereafter referred to as HHP; 
Harris \& Harris 2000, HH00; Harris \& Harris 2002, HH02;
Marleau et al. 2000), the edge-on S0
NGC 3115 (Elson 1997; Kundu \& Whitmore 1998), and 
two dwarf ellipticals in the M81 group (Caldwell et al. 1998). 
Furthermore, Rejkuba et al. (2002) have demonstrated the existence
of a significant intermediate-age AGB population in 
parts of the NGC 5128 halo,
based on deep $U$, $V$, and $K_{s}$ color-magnitude and color-color
diagrams of halo stars resolved by VLT. 

A remarkable result emerging from the {\it HST}/WFPC2 photometry of
the old-halo red giant stars in NGC 5128 is the dominance of
moderately metal-rich stars in the range $-1$ $<$ [m/H] $<$ 0.0
(HHP, HH00, and HH02).  HH00 and HH02  suggested
that the relatively high mean abundance and 
small fraction ($\sim$ 10\%) of metal-poor stars with [m/H] $<$ $-1$
can give strong constraints on the total mass of dwarf galaxies
that could be accreted by NGC 5128
and then disrupted to form its faint stellar halo.

To date, however, there are only a few attempts at physical interpretation
of this material.  HH00 discussed a one-zone model of
chemical evolution of this galaxy and suggested that it
experienced two fairly distinct stages of halo formation: 
an early ``accreting
box'' stage when the galaxy is assembling  through  infall  of star-forming 
gas clumps, and then a major stage of ``closed-box'' evolution 
when the star formation proceeds in the galaxy with little gas infall. 
HH02 extended this basic picture further to a more generalized
accreting-box formation model, in which the infall rate of unenriched gas 
is envisaged to start
at a high rate and then dies away exponentially, while star formation
continues throughout until the gas supply is exhausted.  With appropriate
choices of gas infall rate and the effective chemical yield, excellent
overall matches to the observed MDFs can be obtained.
Recently, Beasley et al. (2002a, b) have further discussed the origin of the stellar
halo MDF within the context of a semi-analytic model of galaxy formation
(GALFORM), based on the Cold Dark Matter (CDM) picture.  In this model, star
formation can take place in two modes, a ``quiescent'' mode within the relatively
unenriched pregalactic clouds, and a ``starburst'' mode whenever two roughly
equal clouds suddenly merge.  They demonstrated that the  
global metallicity distribution within model galaxies comparable to NGC 5128 
in size is broadly consistent with the observed MDF,
although the model is unable to provide any information on the spatial distribution
or radial change in the MDF.

These previous studies are based on one-zone models of chemical evolution 
and thus cannot spatially resolve the various distinct stellar components 
(i.e., halo, bulge,  and disc) in  a galaxy.
Accordingly, comparisons of these models with the MDFs observed at different places
in NGC 5128 have limited value.
%  WE DON'T UNDERSTAND THE FOLLOWING TEXT -- BUT IT DOESN'T SEEM NECESSARY ANYWAY
%by one-zone models with that observed for the {\it stellar halo} of NGC 5128,
%though previous models  provided some insight into the origin of NGC 5128 halo.
%The MDF of the Galactic stellar halo is observed to be remarkably different
%from that of the Galactic stellar disc (e.g., Ryan \& Norris 1991), which 
%implies that we should separately investigate the MDF of a halo and that
%of a  disc (or a spheroidal) 
%so that we can  discuss the origin of stellar haloes of galaxies
%in a more reasonable way, 
%though the difference in MDF between the Galactic stellar halo and disc
%is not necessarily true for other galaxies). 
Numerical simulations which enable us to resolve
spatially the halo, bulge, and disc components and thereby
to investigate the MDF for each of these components should be very helpful
for better understanding the formation of galactic stellar haloes.

The purpose of this paper is to investigate the origin of the MDF 
of the stellar halo of NGC 5128 based on numerical simulations of elliptical
galaxy formation. We here adopt the basic ``merger'' approach (Toomre 1977)
in which elliptical galaxies are proposed to be formed by major merging
of two spiral galaxies.  Within the context of this assumption,
we investigate the final MDF of the outer halo component of the merger remnant 
(i.e., the giant E galaxy) as well as its dependence on the
input parameters of the progenitor discs (e.g., the
MDFs of discs, bulge-to-disc-ratio, and halo mass fraction of the disc).
Comparing the simulated MDFs with the observed ones for NGC 5128,
we discuss the following points: (1) how the dynamics of galaxy merging
is important for the formation of the gE stellar halo,
(2) how the initial MDF of a disc (or bulge) in a merger controls the final
MDF of the stellar halo,
(3) whether the bulge-to-disc ratio of a merger progenitor spiral is important
for the determination of the stellar halo's MDF,
(4) what initial conditions of galaxy mergers in the present simulations
can best give the MDF similar to the observed one,
(5) the relevance of the MDF for the globular cluster system, which is
different from the field-halo stars,
and (6) whether, on the grounds of the merger picture, we should always expect the
MDFs of stellar haloes in elliptical galaxies (E) to be systematically 
different (more metal-rich) from those in  late-type spirals (Sp).
 
The plan of the paper is as follows: In the next section,
we describe our  numerical models  for stellar halo formation
in galaxy mergers. 
In \S 3, we 
present the numerical results
mainly on the final MDFs of  merger remnants (i.e., elliptical galaxies) 
for variously different merger models.
In \S 4, we predict a possible relationship between
MDFs of stellar haloes and physical properties of their host elliptical
galaxies.  
In this section, we also discuss  the origin of M31's metal-rich stellar halo. 
We summarize our  conclusions in \S 5.

\section{The model}

\subsection{Merger model parameters}

In the model calculations, 
we numerically investigate dynamical evolution of major galaxy
mergers between spiral galaxies  whose primary components 
are dark matter haloes,
stellar discs, stellar haloes, and globular clusters (GCs).
In the present study, we do not include gaseous components 
(thus no star formation) in
the simulations so that all of the halo stars in the merger remnant
come from the pre-existing stellar components of the
progenitor spirals.  
The density profiles of dark matter and disc components 
in a spiral follow the
Fall-Efstathiou (1980) model.  
The total mass and the size of a disc are assumed to be
$M_{\rm d}$ and $R_{\rm d}$, respectively. 
All masses and lengths are measured in units
of $M_{\rm d}$ and $R_{\rm d}$ unless otherwise
specified. Velocity and time are measured in units of $v$ = $
(GM_{\rm d}/R_{\rm d})^{1/2}$ and $t_{\rm dyn}$ = $(R_{\rm
d}^{3}/GM_{\rm d})^{1/2}$, respectively, and the gravitational
constant $G$, is assumed to be 1.0. 
If we adopt $M_{\rm d}$ = 6.0 $\times$ $10^{10}$ $ \rm M_{\odot}$
and $R_{\rm d}$ = 17.5 kpc as fiducial values, then
$v$ = 1.21 $\times$ $10^{2}$ km/s and $t_{\rm dyn}$ = 1.41
$\times$ $10^{8}$ yr.

In the disc model, the rotation curve becomes nearly
flat at $R$ = 0.35 (where $R$ is distance from the centre of the
disc) with the maximum rotational velocity $v_{\rm m}$ = 1.8 in
our units (220 km s$^{-1}$).  The corresponding total mass of the
dark matter halo (within 1.5$R_{\rm d}$) is 4.0 in our units.
The velocity dispersion of the dark halo component at a given point is set
to be isotropic and given according to the virial theorem.  The
radial ($R$) and vertical ($Z$) density profiles of the disc are
assumed to be proportional to $\exp (-R/R_{0}) $ with scale
length $R_{0}$ = 0.2 and to ${\rm sech}^2 (Z/Z_{0})$ with scale
length $Z_{0}$ = 0.04 in our units, respectively.  
In addition to the
rotational velocity caused by the gravitational field of disc,
bulge, and dark halo components, the initial radial and azimuthal
velocity dispersions are assigned to the disc component according to
the epicyclic theory with Toomre's parameter $Q$ = 1.5.  The
vertical velocity dispersion at given radius is set to be 0.5
times as large as the radial velocity dispersion at that point,
as is consistent with the observed trend of the Milky Way (e.g.,
Wielen 1977).

The density profile of the central
bulge with a mass of $M_{\rm b}$ and effective radius of $R_{\rm e}$ 
(in our units) is
represented by the de Vaucouleurs $R^{1/4}$ law.
We mainly investigate two different bulge models:
(a) a model with $M_{\rm b}$ = 0.2 (in our units, corresponding to  
1.2  $\times$ $10^{10}$ $ \rm M_{\odot}$) and $R_{\rm e}$ =  0.04
(in our units corresponding to 0.7kpc) and (b) a model with
$M_{\rm b}$ = 0.6 (3.6 $\times$ $10^{10}$ $ \rm M_{\odot}$) 
and 0.11 (2kpc). These two cases are intended to correspond roughly
to the Milky Way and to M31, respectively.
Each spiral is assumed to have  an extended stellar halo with a 
density distribution of $\rho (r) \sim r^{-3.5}$,
where $r$ is the distance from the center of the spiral.
The mass fraction of the stellar halo is represented by $f_{\rm h}$
and considered to be a free parameter in the present study. 

We also include a population of collisionless  particles to represent
a part of the globular cluster system in each galaxy.
The Milky Way and M31 are observed to have 160 $\pm$ 20 and 400
$\pm$ 55 globular clusters, respectively (van den Bergh 1999).
We therefore assume each spiral to have 200
old, metal-poor globular clusters with a number density
distribution like that of the Milky Way
system (i.e., $\rho (r) \sim r^{-3.5}$). 

In total, the numbers of 
particles used for modeling the merger
are 20000 for the dark matter haloes, 20000 for the discs, 12000 (4000) for 
bulges (The Galactic bulge),
4000 for the stellar haloes, and 400 for GCs.

The orbits of the two spirals in any given run of the model
are set to be initially in the $xy$ plane by definition, and the
distance between the centre of mass of the two spirals ($r_{\rm
in}$) is 6$R_{\rm d}$ (105 kpc). 
The pericentre distance ($r_{\rm p}$) and the orbital
eccentricity ($e_{\rm p}$) are assumed to be free parameters
which control the orbital angular momentum and energy of the merging
galaxy.  We show here the particular cases for $e_{\rm p}$ =1 and
$r_{\rm p}$ = $R_{\rm d}$ (17.5 kpc), because 
the dynamics of halo formation do not
depend strongly on these merger parameters.
The spin of each spiral in a merging
galaxy is specified by two angles $\theta_{i}$ and
$\phi_{i}$ (in units of degrees), where the suffix $i$ is used to
identify each galaxy.  Here, $\theta_{i}$ is the angle between
the $z$ axis and the vector of the angular momentum of the disc,
and $\phi_{i}$ is the azimuthal angle measured from $x$ axis to
the projection of the angular momentum vector of the disc onto
the $xy$ plane. We specifically investigate the following three 
models with different disc inclinations with respect to the
orbital plane: (a) a prograde-prograde model represented by ``PP''
with $\theta_{1}$ = 0, $\theta_{2}$ = 30, $\phi_{1}$ = 0, and
$\phi_{2}$ = 0, (b) a prograde-retrograde (``PR'') with $\theta_{1}$
= 30, $\theta_{2}$ = 240, $\phi_{1}$ = 90, and $\phi_{2}$ = 0, (c) a
retrograde-retrograde (``RR'') with $\theta_{1}$ = 180,
$\theta_{2}$ = 210, $\phi_{1}$ = 0, and $\phi_{2}$ = 0.
All calculations related to the above collisionless evolution
have been carried out on the GRAPE board (Sugimoto et al. 1990) 
at the Astronomical Institute of Tohoku University.

\subsection{MDFs of spirals}

A key point of these simulations is to trace out what generates the
MDF of the merger product halo.
We thus assume that each stellar component within the progenitor spirals
has its own characteristic MDF which does not
change during a simulation of a galaxy merger.
The initial MDFs for each of the stellar components (i.e., bulge, disc, halo, and GC)
in a merger progenitor spiral are assumed to be different from one another,
and are defined to mimic the observed ones in the Milky Way.

\subsubsection{Disk}
Recent photometric and spectroscopic observations of the Galactic 
stellar and gaseous components and gaseous components of external galaxies 
have demonstrated that disc galaxies have metallicities that depend on
location within the disc, i.e., metallicity gradients
(e.g., Zaritsky, Kennicutt, and Huchra 1994; Friel 1995).
Considering these observations, 
we allocate metallicity to each disc star according to its initial position: 
at $r$ = $R$, 
where $r$ ($R$) is the projected distance (in units of kpc) 
from the center of the disc, the metallicity of the star is given as: 
\begin{equation}
{\rm [m/H]}_{\rm r=R} = {\rm [m/H]}_{\rm d, r=0} + {\alpha}_{\rm d} \times {\rm R}. \;
\end{equation}  
We adopt two plausible values for the slope ${\alpha}_{\rm d}$:  (a) $-0.091$
from Friel (1995) in which the Galactic stellar metallicity gradient
is estimated from Galactic open clusters, and
(b) $-0.056$ from Zaritsky et al. (1994) in which
{\it gaseous} metallicity gradients of disc galaxies are estimated
from HII regions of these.   (It should be  noted here that we do not have
any observational results which have {\it directly} measured
{\it stellar} metallicity gradients of external disc galaxies.)
The models are labeled as ``OC''-type (those with the Friel et al. gradient
defined by open cluster metallicities) and ``GAS''-type (those with the
Zaritsky et al. gradient based on nebular and gaseous abundances).
The value of the zeropoint ${\rm [m/H]}_{\rm d, r=0}$ is chosen such that 
the mean metallicity of a disc (represented by ${\rm [m/H]}_{\rm d}$
 $=-0.15$ for the fiducial model) 
can be different for different models. 

We comment that the steeper ``OC''
slope might {\sl a priori} be considered the preferable one, since it is
based on the old- to intermediate-age stars which represent the bulk of
the stars in the disc, and these stars
are likely to be most relevant to mergers at the earlier epochs we are 
attempting to model here.  By contrast, the ``GAS'' slope represents the
youngest stars and present-day gas, which are a much smaller fraction of the
disc mass.  As will be seen below, $\alpha_{\rm d}$ turns out to be perhaps
the single most important parameter for understanding the outcome of the
post-merger halo.

\subsubsection{Bulge}

The Galactic bulge is also observed to have a metallicity gradient
(e.g., McWilliam \& Rich 1994; Wyse et al. 1997; Frogel et al. 1999).
Therefore in the models we assign the metallicity of a bulge star at $r$ = $R$, 
where $r$ ($R$) is the projected distance (in units of kpc) of the star 
from the center of the bulge, to be:
\begin{equation}
{\rm [m/H]}_{\rm b, r=R} = {\rm [m/H]}_{\rm b, r=0} + {\alpha}_{\rm b} \times {\rm R}. \;
\end{equation}  
We adopt the metallicity gradient from Frogel et al. (2000) in which
the slope ${\alpha}_{\rm b} = -0.43$.
For the model with $M_{\rm b}$ = 0.2, we use  
${\rm [m/H]}_{\rm b, r=0} = 0.034$.
Since there is an observed relation between luminosity ($L$) and metallicity ($Z$)
for spheroidal galaxies ($L$ $\propto$ $Z^{0.4}$; Mould 1984),
we allocate a larger metallicity of ${\rm [m/H]}_{\rm b, r=0}$ 
to the model with a larger bulge.  
For example, the value of ${\rm [m/H]}_{\rm b, r=0}$ in the model 
with $M_{\rm b}$ = 0.6 is set to be 0.19.

\subsubsection{Halo and GCs}

There is no clear evidence of radial metallicity gradients  
in the Galactic halo stars or halo globular
cluster system (e.g., Searle \& Zinn 1978; Chiba \& Yoshii 1998).
We accordingly do not allocate a metallicity to each halo star (GC)
based on its initial position. Instead, we assume that 
the halo stars (GCs) in a model have a Gaussian metallicity distribution with a
mean of ${\rm [m/H]}_{\rm h}$ (${\rm [m/H]}_{\rm gc}$) $= -1.4$
and dispersion of $\sigma$([m/H]) = 0.3 (and truncated at $4\sigma$). 
The adopted value of $\sigma$([m/H]) = 0.3 has already been demonstrated
to be a reasonable value in explaining the observed MDF of the Galactic stellar
halo and GCs (e.g., C\^ot\'e et al. 2000). 
We set the value of ${\rm [m/H]}_{\rm gc}$ to be $-1.6$ for all models. 

Most large galaxies also have GCs which
are more metal-rich and usually more concentrated to the galaxy center
than the metal-poor halo clusters.  In the Milky Way and M31, these
more metal-rich clusters (which have a mean [m/H] $\simeq -0.5$ and higher
overall rotation speed around the galaxy center) form about a quarter to
a third of the whole GC population in both M31 and the
Milky Way, and can plausibly be interpreted as
a ``bulge'' component (see Perrett et al. 2002; Harris 2001; Minniti 1996).
The MDF for the entire globular cluster system in these and many other
large galaxies can usually
be described as having a bimodal form (e.g. Larsen et al. 2001), often with
a clear separation between the halo and bulge components.  
We do not explicitly include any bulge GCs in the simulations, because 
(as will be seen below) they behave similarly to the bulge stars during
the merger, and at this
stage we are interested primarily in the formation of the
elliptical halo.  We will discuss the GC results further in \S 3.2.

The MDFs for the various components of a progenitor spiral 
are given in Fig. 1
for the fiducial model (described in detail later) with
the ``OC'' type metallicity gradient,   ${\rm [m/H]}_{\rm d}$ = $-0.15$,
$M_{\rm b}$ = 0.2 (The Galactic bulge model), and ${\rm [m/H]}_{\rm h}$ = $-1.4$.

\begin{figure}
%\vspace{11pc}
%\psfig{file=f1.ps}
%\psfig{file=f1.ps}
%\psfig{file=f1.eps,width=18.5cm}
\psfig{file=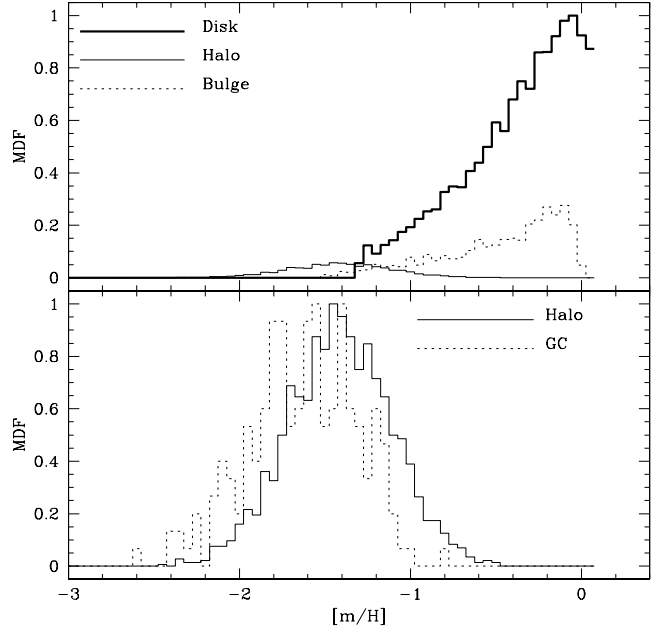,width=8.5cm}
\caption{ 
{\it  Upper:} The MDFs of each of stellar component, disc (thick solid),
halo (thin solid), and bulge (dotted) in a spiral galaxy for the fiducial model M1. 
``OC''-type metallicity gradient and a bulge mass ($M_{\rm b}$ = 0.2) 
are assumed in this model.
{\it  Lower:} The MDFs of  the stellar halo (solid)
and globular clusters (represented by GC, dotted) in a spiral galaxy 
for the fiducial model M1. 
Here ${\rm [m/H]}_{h}$ = $-1.4$ and  ${\rm [m/H]}_{gc}$ = $-1.6$ 
 are assumed in this model.
}
\label{Figure. 1}
\end{figure}
\begin{figure}
%\vspace{11pc}
%\psfig{file=f1.ps}
%\psfig{file=f1.ps}
%\psfig{file=f1.eps,width=18.5cm}
\psfig{file=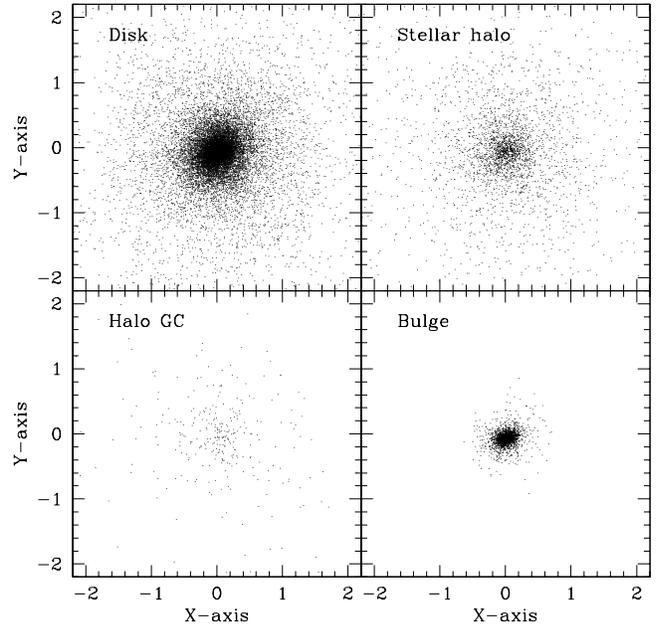,width=8.5cm}
\caption{ 
Final mass distribution of stellar components that 
are initially disc stars (upper left),
halo ones (upper right), GCs (lower left), and bulge stars (lower right)
in the elliptical galaxy formed by major merging in the fiducial model M1 
at $T$ = 20 in our units (2.8 Gyr). One frame measures 77 kpc on  a side. 
Note that the halo region ($R$ $>$ 0.5) of this elliptical galaxy is dominated
by disc stars.  
}
\label{Figure. 2}
\end{figure}

\begin{figure}
%\vspace{11pc}
%\psfig{file=f1.ps}
%\psfig{file=f1.ps}
%\psfig{file=f1.eps,width=18.5cm}
\psfig{file=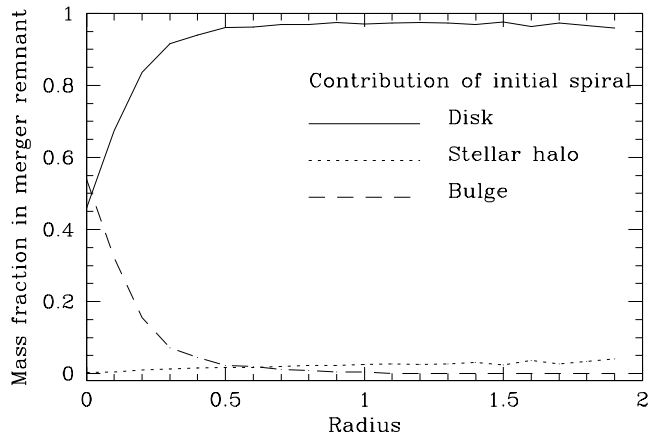,width=8.5cm}
\caption{ 
Radial dependence of mass fraction of each stellar component,
disc  (solid), halo (dotted), and bulge stars (dashed) in the merger remnant 
for the fiducial model M1. Note that more than 90 \% of stars with
$R$ $>$ 0.5 (regarded as ``stellar halo'' of the elliptical in
the present study) are initially stars in the disc of the merger progenitor
spirals.
}
\label{Figure. 3}
\end{figure}

\subsection{Main points of analysis}

For the NGC 5128 halo, MDFs have been obtained at projected distances of
8 kpc, 21 kpc, and 31 kpc from the galaxy center
(HHP, HH00, HH02).  The two outer-halo locations (21 and 31 kpc) turned
out to give identical results for the MDF.  For comparison with the
models, we investigate the model MDFs of two halo regions which correspond
closely with the these inner and outer regions, one at
0.5 $<$ $R$ $\le$ 1.0 (referred to as
``inner halo'') and the other at 1.0 $<$ $R$ $\le$ 2.0 (``outer halo'') in 
each merger model. 
The simulated MDF of the ``inner halo'' and that of the ``outer halo''
are compared with the observed MDF derived for the 8 kpc field  
and with that derived for the combined 21 and 31 kpc fields. 
Although we investigated $\sim 20$ models,  
we here present only the results of {\it the models which  are not obviously
inconsistent with observational results}.  That is, we attempt to outline
the range of model parameter space which leads to realistic results.
For example, our simulations have shown that,
if the {\sl halo mass fraction} of the merger progenitor
spirals is greater than 0.2,
the resultant MDF has far too many metal-poor stars to agree with
the observed MDF. Equally, if the initial disc's mean 
metallicity in a merger is much larger than [m/H] = 0.0,
the peak value of the resultant MDF is too  metal-rich to be consistent
with observations.
We will not show the results of these obviously inconsistent models  
here but, rather, we present the results of the more realistic models 
in which the mean disc metallicity (${\rm [m/H]}_{\rm d}$ ) 
is equal to -0.15
and 0.0, and in which the halo mass fraction is no more than 0.2.

Below, we describe the results of 10 models and in Table 1 summarize 
their input parameters: model number (column 1), 
bulge mass fraction $M_{\rm b}$ (column 2),
metallicity gradient type (3),
mean metallicity of disc in  ${\rm [m/H]}_{\rm d}$  (4),
orbital type of galaxy merging (5),
halo mass fraction $f_{\rm h}$ (6),
and initial halo mean metallicity ${\rm [m/H]}_{\rm h}$ (7). 
For convenience we refer to the model M1 as the {\sl fiducial model}.
It shows typical behavior of stellar halo formation in major merging
and its MDF is reasonably consistent with the observations
 of the NGC 5128 stellar halo.
Accordingly we describe mainly the results of this model in the following sections.
The MDFs of this model for each component (disc, bulge, halo, and GCs)
 are summarized in Table 1.
The metallicity gradient types ``OC'' and ``GAS'' in the third column
represent the two choices for disc metallicity gradient described above.
Orbital type labels ``PP'', ``PR'' and ``RR'' represent
a nearly prograde-prograde orbital configuration, a prograde-retrograde one,
and a nearly retrograde-retrograde one. The structural properties of merger remnants
are briefly summarized in Appendix A.

\begin{figure}
%\vspace{11pc}
%\psfig{file=f1.ps}
%\psfig{file=f1.ps}
%\psfig{file=f1.eps,width=18.5cm}
\psfig{file=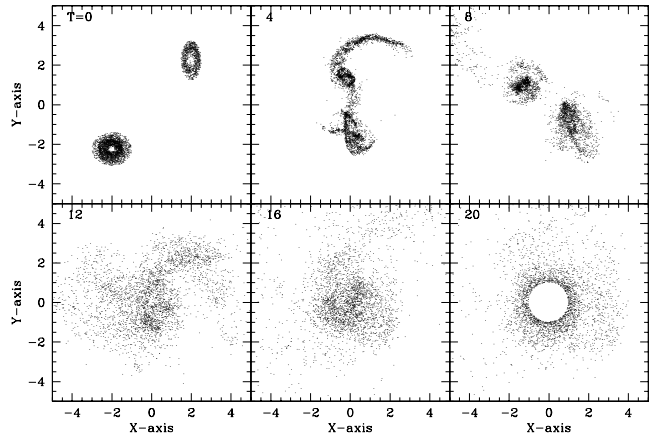,width=8.5cm}
\caption{ 
Morphological evolution of disc stars that become halo stars
with $R$ $>$ 1 (where $R$ is the projected distance 
from the center of the merger in our units)  at $T$ = 20 (2.8 Gyr)
in the fiducial merger model M1 projected onto the  $x$-$y$ plane.
The scale is given in our units, and accordingly each frame measures
175 kpc on a side. Note that the outer halo stars of the merger remnant 
are initially located in the outer part of the progenitor discs.
Note also that tidal stripping (or angular momentum redistribution) of  disc stars 
during major galaxy merging is a key physical process of the stellar halo formation
in the elliptical galaxy.  The six boxes show this evolution at times of
T=0,4,8,12,16,and 20.
}
\label{Figure. 4}
\end{figure}

\section{Results}

\subsection{Dynamics of stellar halo formation and MDFs of ellipticals}

Figs. 2 and 3 summarize the final mass distributions of the stellar components 
that are initially within discs, haloes, and bulges of the merger progenitor spirals
for  the fiducial major merger model M1. Since the bulges initially have more  
compact configurations than the discs, only a small fraction of bulge stars
are tidally stripped to form the inner stellar halo 
with 0.5 $<$ $R$ $\le$ 1.0 of the merger remnant,
where $R$ is  the radius from the center of the developed elliptical
galaxy in units of $R_d$. 
The stars in the discs, on the other hand, are very efficiently stripped
owing to the strong tidal field of major merging (or by angular momentum 
redistribution during merging) so that they end up in both the
inner and outer halo 
(1.0 $<$ $R$ $\le$ 2.0) of the elliptical galaxy.
Although the initial stellar haloes of the spirals 
are also stripped (or `puffed up') efficiently
to form the inner and the outer stellar haloes, {\sl the stellar halo
of the resulting elliptical is dominated by the stars that were 
initially within the progenitor discs}; about 95\% of 
the elliptical halo comes from the initial disc stars at 1 $<$ $R$
(For comparison, the outer parts of the spirals with 1 $<$ $R$ $\le$ 2
are by definition dominated {\it only} by their metal-poor halo stars). 
These results imply that major mergers can efficiently populate the
stellar haloes of their product elliptical galaxies  
by disc stars from within their progenitor spirals.

We note also from Fig. 4 that the (outer) stellar halo of the elliptical galaxy 
comes from the outer part of the merger progenitor disc,
because the tidal stripping (or angular momentum redistribution)
is more efficient for the outer part of the disc than for the inner part.
Furthermore, Fig. 4 shows that, although the  violent dynamical relaxation of major
merging results in strong mixing of stars of the two discs, 
we can clearly see the inhomogeneous distribution of halo stars at 2 $<$ $R$
(e.g., the upper right part of the halo appears to be denser than the lower left). 
This implies that if we can observe {\it the global}  stellar halo or
surface intensity distribution
of an elliptical galaxy, it may show a certain degree of inhomogeneity.
%  HOW LONG DOES THE INHOMOGENEITY LAST?
%  I.E. AFTER A FEW GYR DOES IT DIE AWAY?  WOULD ONLY RECENT MERGERS
%  SHOW THE EFFECT?  (AND FOR RECENT MERGERS, MAYBE MORE OBVIOUS
%  TRACES WOULD BE IN THE CENTER BULGE REGION?)

As is shown in Fig. 5, the surface density of the merger halo is 
greater by roughly 2 orders of magnitude at a given radius
than the initial  spiral halo
owing to the stripped disc stars.  
This result suggests that the surface brightness of stellar haloes 
in  elliptical galaxies will be systematically higher than for haloes in spiral
galaxies if most ellipticals are formed by major merging.
However, the slope of the halo light distribution does not change 
significantly compared with that of the initial spiral halo.

\begin{figure}
%\vspace{11pc}
%\psfig{file=f1.ps}
%\psfig{file=f1.ps}
%\psfig{file=f1.eps,width=18.5cm}
\psfig{file=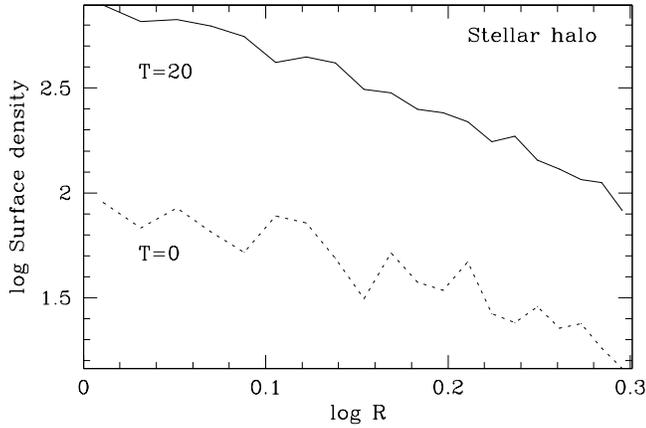,width=8.5cm}
\caption{ 
Radial distribution of the surface (number) density of stars located
in the halo region  
of the merger model M1 at $T$ = 20 (solid)
and those in the halo 
region of the progenitor spiral at $T$ = 0 (dotted).
Here the results are given for the ``outer'' halo region 
(1 $<$ $R$ $\le$ 2). 
The scale is given in our units, and accordingly the outer halo
region corresponds to 17.5 $<$ $R$ $\le$ 35.0 kpc. 
}
\label{Figure. 5}
\end{figure}

\begin{figure}
%\vspace{11pc}
%\psfig{file=f1.ps}
%\psfig{file=f1.ps}
%\psfig{file=f1.eps,width=18.5cm}
\psfig{file=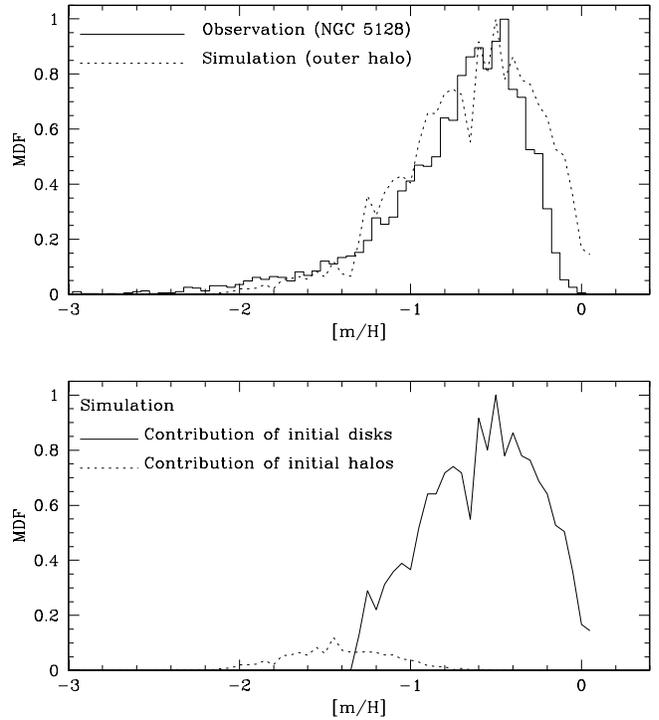,width=8.5cm}
\caption{ 
{\it Upper:} MDFs derived for observations (solid) and the simulation of model M1 (dotted).
The MDF for observations is constructed from halo stars with the projected distance
of 21 and 31 kpc in NGC 5128 halo fields. The MDF for the simulation
is constructed from the simulated stars with 1 $<$ $R$ $\le$ 2 
(17.5 kpc $<$ $R$ $\le$ 35 kpc),  where $R$ is 
the distance of a star from the center of the merger remnant.
Note that the simulated MDF shows a peak around [m/H] $\sim$ $-0.4$,
which is similar to the observed peak.
{\it Lower:} MDFs for stars that are initially disc stars (solid)
and halo ones (dotted) in  the simulation of model M1.
These MDFs are also estimated for the outer halo region with 1 $<$ $R$ $\le$ 2.
Bulge stars do not contribute to the MDF in this outer halo at all.
}
\label{Figure. 6}
\end{figure}

\begin{figure}
%\vspace{11pc}
%\psfig{file=f1.ps}
%\psfig{file=f1.ps}
%\psfig{file=f1.eps,width=18.5cm}
\psfig{file=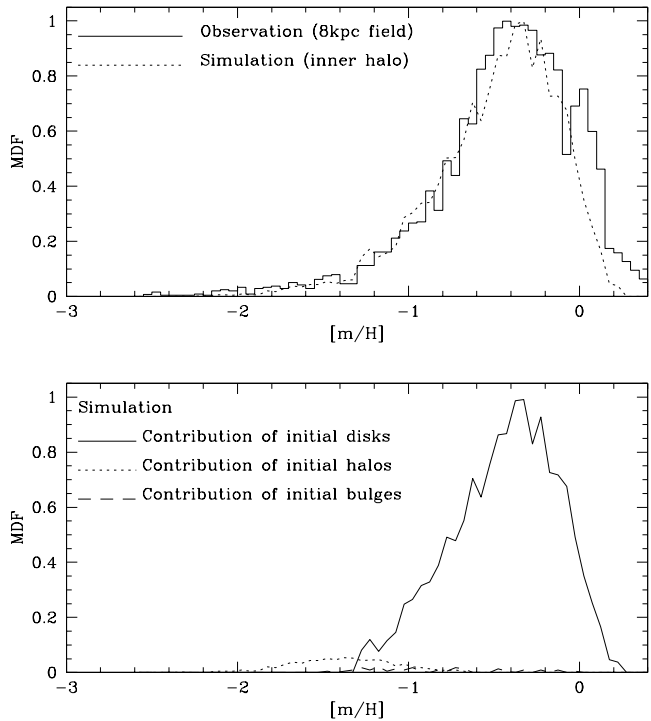,width=8.5cm}
\caption{ 
The same as the Figure 6 but for the inner halo region
with 0.5 $<$ $R$ $\le$ 1.0.
The MDF for observations is constructed from halo stars with the projected distance
of 8  kpc in NGC 5128 halo fields. 
In the lower panel, MDFs for stars that are initially disc stars (solid)
and halo ones (dotted), and bulge ones (dashed) 
are plotted.
}
\label{Figure. 7}
\end{figure}
   
A significant result of the halo formation by stripping of disc stars
is that the MDF of the stellar halo of the elliptical galaxy 
becomes remarkably different from that of the progenitor spirals.
We show this, and the comparison with the data, in Figures 6 and 7.
%The model MDF and the observations both have peak values of [m/H] $\sim$ $-0.4$ 
%and thinly populated metal-poor tails ([m/H] $<$ $-1.4$).
%The derived peak value is significantly higher than that ($\sim$ $-1.6$) of the
%MDF in the Galactic stellar halo (e.g., Ryan \& Norris 1991),
%which clearly suggests that major merging can be responsible 
%for the observed difference in the peak values of MDFs between
%the Galaxy and bulge-dominated M31 and elliptical galaxy NGC 5128.
By comparing Fig.1 and Fig. 6, we also note that
the difference in MDFs between the halo and the disc (spheroidal)
components in this model  becomes much smaller after merging.
This implies that the MDFs of the stellar haloes in elliptical galaxies
are  less different from those of the main spheroidal bodies than they are  in spirals.

As is shown in the lower panel of Fig. 6, the stellar halo of the
emergent E galaxy is a genuine composite population in this formation scheme.
The relatively metal-rich part which is the vast majority 
([m/H] $>$ $-1$) come from the progenitor discs,
 whereas the thin metal-poor part comes from the progenitor haloes. 
A {\sl very} small fraction at the metal-rich end is from the original
bulges.  

Fig. 7, showing the inner halo (0.5 $<$ R $\le$ 1.0) of the elliptical
galaxy, has an MDF which is shifted even further to the metal-rich end,
with a broad peak at [m/H] $\sim -0.4$ and an extension to well above
Solar metallicity, and an even sparser low-metallicity tail than the outer halo.
Here, a higher fraction of the stars are from further in within the
progenitor discs ($\sim$ 97 \%), as well as from the initial bulges ($\sim$ 2 \%).
The simulated MDF is not as broad as the observed inner halo MDF and, in
particular, does not extend as far into the metal-rich regime.  This may
be significant, particularly since the observed data are likely to be 
more incomplete for the reddest stars (HH02).   
Thus the principal driver of the observed metallicity gradient in the 
resulting E galaxy halo is the presence of the initial 
metallicity gradients in the disc.
Fig. 8  demonstrates  that the MDF in the very outer part of the halo
($R$ $>$ 2, corresponding to $R$ $>$ 35 kpc) shows a much wider peak around
$-1$ $<$ [m/H] $<$ $-0.4$ and a larger fraction of 
metal-poor ([m/H] $<$ $-1.0$) stars 
compared with the MDFs for the inner and the outer
haloes.  This is mainly because the very outer  halo with $R$ $>$ 2 
consists of more metal-poor stars coming from the extreme outer parts of the
merging discs. The predicted  difference in the MDFs between the outermost halo 
(2 $<$ $R$) and the already  observed inner and outer haloes 
(i.e., 0.5 $<$ R $\le$ 1.0 and  1.0 $<$ R $\le$ 2.0)
can be tested by future observations on the MDF of the NGC 5128 stellar halo
at  projected distances of larger than 30 kpc.

\begin{figure}
%\vspace{11pc}
%\psfig{file=f1.ps}
%\psfig{file=f1.ps}
%\psfig{file=f1.eps,width=18.5cm}
\psfig{file=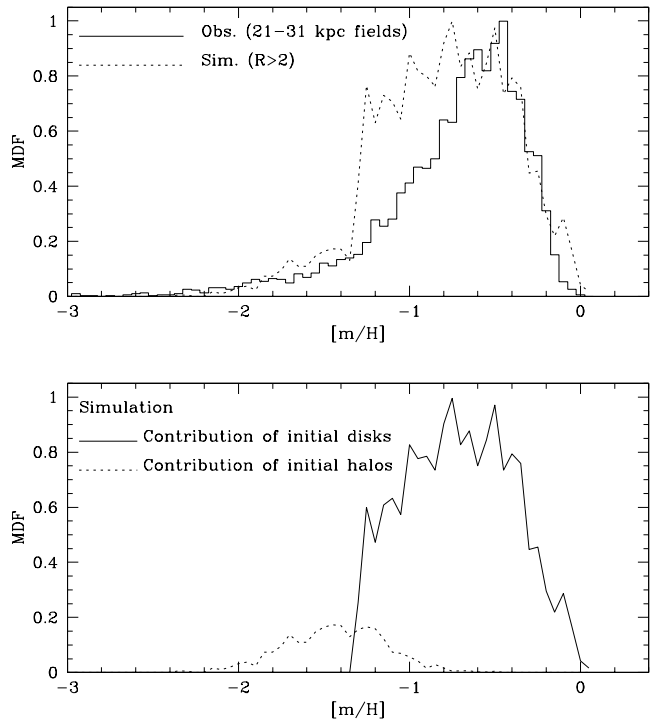,width=8.5cm}
\caption{ 
The same as the Figure 6 but for the
very outer halo regions with 2.0 $<$ $R$.
The MDF for observations is constructed from halo stars with the projected distance
of 21 and 31 kpc in NGC 5128 halo fields. 
}
\label{Figure. 8}
\end{figure}

\begin{figure}
%\vspace{11pc}
%\psfig{file=f1.ps}
%\psfig{file=f1.ps}
%\psfig{file=f1.eps,width=18.5cm}
\psfig{file=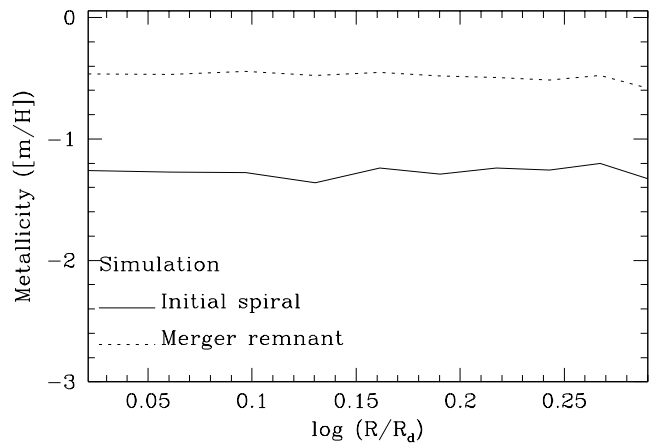,width=8.5cm}
\caption{ 
Metallicity gradient of the stellar halo with 1 $<$ $R$ $\le$ 2) 
in the initial spiral (solid) and the merger remnant (dotted) for the model M1.
Note that values of log ($R/R_d$) = 0.08 and 0.25 correspond to
 the approximate locations 
of the observed 21kpc and 31 kpc fields. 
}
\label{Figure. 9}
\end{figure}

\begin{figure}
%\vspace{11pc}
%\psfig{file=f1.ps}
%\psfig{file=f1.ps}
%\psfig{file=f1.eps,width=18.5cm}
\psfig{file=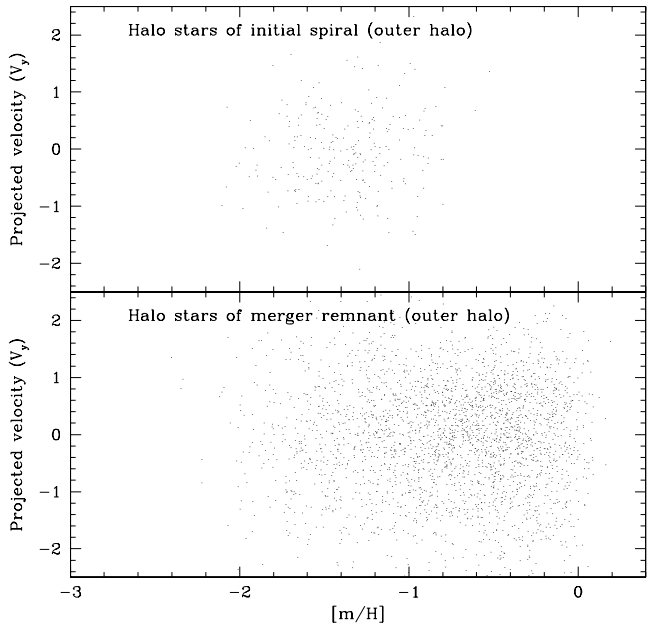,width=8.5cm}
\caption{ 
Distribution of halo stars  with 1 $<$ $R$ $\le$ 2 (outer halo)
on $V_{\rm y}$-[m/H] diagram for the merger progenitor spiral (upper)
and the merger remnant (lower) in the model M1. 
$V_{\rm y}$ here represents the velocity
of a star projected onto $y$-axis of the simulation.
Note that there are a larger number of halo stars with [m/H] $>$ $-0.6$
and the absolute magnitude of $V_{\rm y}$ larger than 1.8 (250 km s$^{-1}$)
only in the merger remnant.
A physical interpretation of this is given in the text. 
}
\label{Figure. 10}
\end{figure}

\begin{figure}
%\vspace{11pc}
%\psfig{file=f1.ps}
%\psfig{file=f1.ps}
%\psfig{file=f1.eps,width=18.5cm}
\psfig{file=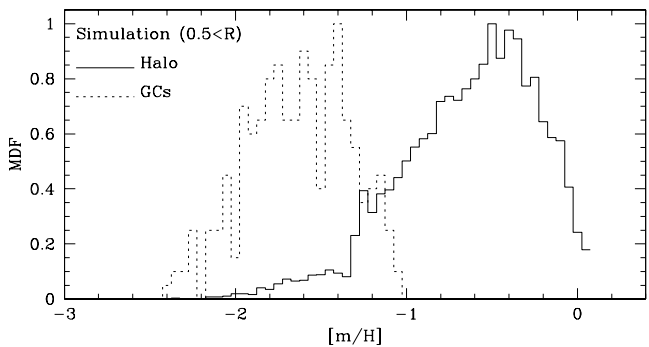,width=8.5cm}
\caption{ 
MDFs of halo stars (solid) and halo GCs (dotted) in 
the merger remnant of the model M1 at $T$ = 20. 
The MDFs are constructed from the stellar components with 0.5 $<$ $R$ 
(i.e., both the inner and the outer stellar components). 
Note that there is a clear difference in MDFs between the halo and the GC.
This clear different is not seen in the initial spiral's MDF (See Fig. 1),
which suggests that dynamical processes of galaxy merging cause
such a remarkable difference in the MDFs between the halo and the GCs
in the model M1. 
}
\label{Figure. 11}
\end{figure}

In Fig. 9, we show the dependence of mean metallicity on galactocentric
distance for the merged elliptical and the initial spirals. 
The {\sl outer} stellar halo of the model elliptical 
(1 $<$ $R$ $\le$ 2) does not show any important
metallicity gradient, though its progenitor {\it discs} have 
negative metallicity gradients. This is mainly  because 
violent dynamical relaxation during galaxy merging
can cause efficient mixing of stellar populations (with different metallicity)
initially located in different regions of the discs, and this 
almost completely smoothes out  
the initial disc's metallicity gradient.
(It should be noted here that if we observed the gradient {\it throughout}
the elliptical, we can still see the negative gradient, which becomes steeper
farther in to the center.) 
Fig. 9 also demonstrates, as indicated above, that the mean metallicity of the stellar  halo
of the elliptical  is  much higher ($\sim$  a factor of 6) 
than that in the progenitor spiral
{\it at all points} in the outer halo regions.
This is remarkably consistent with the HST data for the NGC 5128
fields at 21 kpc and 31 kpc whose MDFs are virtually indistinguishable (HH00).
We suggest that the near-zero metallicity gradient of the outer stellar halo
is an observational test which can assess the validity of
the merger scenario of  elliptical galaxy formation.

In Fig. 10 we show the distribution of the halo stars
on the $V_{\rm y}$-[m/H] diagram (radial velocity vs.~metallicity)
for the progenitor spirals and the elliptical. Only the elliptical 
halo contains stars with both high projected
(radial) velocity ($V_{\rm y}$ $\sim $ 2 in our units, corresponding to 
$\sim$ 280 km s$^{-1}$) and high metallicity ([m/H] $>$ $-0.6$).
Some fraction of these outer halo stars with rather high projected velocity 
also have large tangential velocities and thus high 
angular momentum with  respect to the elliptical.
The origin of these halo stars with high angular momentum is
closely associated with angular momentum transfer, from inside to outside,
during major galaxy merging; orbital angular momentum of the merging spirals
is converted into the intrinsic angular momentum of such outer halo stars. 
Based on kinematics and spatial distribution of the planetary nebula (PN) system
in NGC 5128,
Hui et al. (1995) found a possible evidence of the PN's rotation increasing
with radius and discussed the origin of this elliptical galaxy.
Additional new velocity analyses are given by Peng et al. (2002).
Although the necessary spectroscopic data to obtain kinematical
properties of individual halo stars in NGC 5128 are beyond current
observational capabilities, the results in Fig. 10 suggest that,
future such observations may
find a larger fraction of  relatively metal-rich outer-halo
stars with high angular momentum in NGC 5128.

\subsection{Globular clusters}

By comparing  Fig. 1 with Fig. 11, 
we can clearly see that the difference between the GC and field-star MDFs
in this merger model becomes much more conspicuous after elliptical galaxy formation.
This is essentially because the GCs start out as part of a ``hot'', pressure-supported
component of the spiral haloes and so behave differently from the disc stars during
the merger.  The MDF of the halo clusters thus does not change markedly
during merging, whereas that of the halo can be dramatically changed  
owing to the addition of large numbers of stripped disc stars.  In the product elliptical,
we therefore end up with metal-poor
clusters embedded in a moderately metal-rich and massive halo.

This dynamically constructed difference in MDFs would not be 
changed if we were to include gas and merger-induced formation 
of metal-rich GCs, since numerical simulations have already
demonstrated that these would mostly be
located in the central region of the merger remnant and therefore would
not contribute to the MDF of GCs in the outer halo region (Bekki et al. 2002).
In addition, the ``bulge'' GCs which were already present in the progenitor
spirals (such as are present in both the Milky Way and M31) are more concentrated
to their parent galaxy centers and would stay
in the inner regions along with the bulge stars.
The significant difference in MDFs between the  halo stars and the GCs 
that we note in the M1 simulation
has  been actually observed for M31 (Durrell et al. 2001) 
and NGC 5128 (HHP, HH01, HH02),
both of which have large central spheroidal components.
A mean offset in integrated color (and by hypothesis metallicity) between
GCs and field stars is also classically observed in many giant ellipticals
(e.g. Harris 1991, Brodie \& Huchra 1991).
It has frequently been suggested that the redder, more metal-rich
clusters are similar to the bulk of the field halo and bulge stars, while
the bluer, more metal-poor ones have earlier and different origins, such
as within spiral haloes or dwarf galaxies
(e.g.~ Larsen et al. 2001; Forbes, Brodie, \& Grillmair 1997; Geisler, Lee, \& Kim
1996; C\^ot\'e, Marzke, \& West 1998).
The Milky Way, an Sbc-type spiral with only a small bulge,
is not observed to have such a difference between halo stars and clusters.
We therefore suggest that the observed difference in MDFs between 
stellar haloes and GCs
in elliptical galaxies or early-type spirals with big bulges 
is consistent with the view that their large spheroidal components were formed
by past major merger events. We discuss this point further in \S 4.

\begin{figure}
%\vspace{11pc}
%\psfig{file=f1.ps}
%\psfig{file=f1.ps}
%\psfig{file=f1.eps,width=18.5cm}
\psfig{file=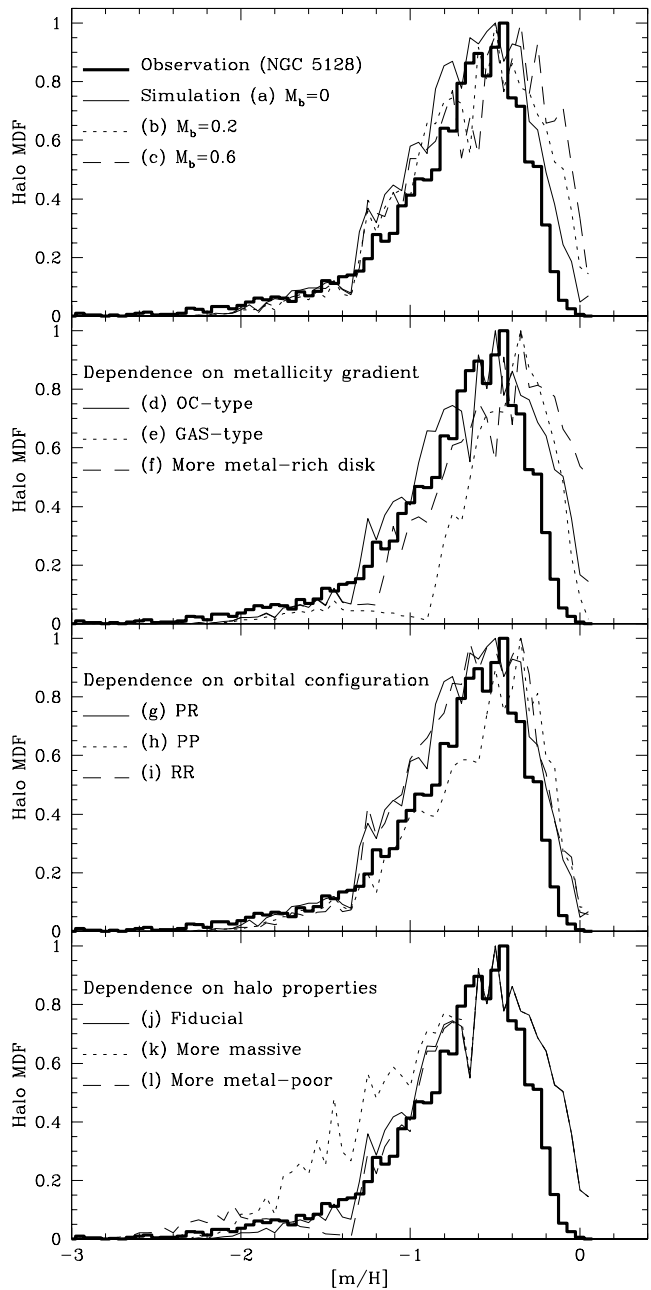,width=8.5cm}
\caption{
Dependence of the simulated MDFs on the initial mass fraction
of bulge (top), 
the MDFs of the merger progenitor discs (second from the top),
orbital configurations of major merging (second from the bottom),
and MDFs of the stellar haloes of the merger progenitor spirals (bottom).
For comparison, the MDF observed for the stellar halo regions  with projected
distances of 21 and 31 kpc in NGC 5128 (HHP, HH00, and HH02) is given by 
a thick solid line 
in each panel.
The results are given for (a) the model with $M_{\rm b}$ = 0 (solid, M3),
(b) $M_{\rm b}$ = 0.2 (dotted, M1), (c) $M_{\rm b}$ = 0.6 (dashed, M2),
(d) ``OC''-type metallicity gradient (solid, M1), 
(e) ``Gas''-type one (dotted, M4),
(f) more metal-rich disc with ${\rm [m/H]}_{\rm d}$ = 0 (dashed, M5),
(g) ``PR'' (prograde-retrograde) orbital configuration (solid, M6), 
(h) ``PP'' (prograde-prograde) orbital configuration (dotted, M7), 
(i) ``RR'' (retrograde-retrograde) orbital configuration (dashed, M8), 
(j) the fiducial set of parameters (solid, M1),
(k) more massive ($f_{\rm h}$ = 0.2) stellar halo (dotted, M9),
and (l) more metal-poor (${\rm [m/H]}_{\rm h}$ = $-2$) stellar halo (dashed, M10). 
}
\label{Figure. 12}
\end{figure}

\subsection{Parameter dependences}

Although the models M1 -- M10 resemble each other in their
dynamics of stellar halo formation and the peak value of the MDFs 
of the resulting stellar haloes, the shapes of the MDFs
depend  on the progenitor
disc metallicity gradient, bulge-to-disc-ratio, and the mass fraction
and the MDFs  of the initial haloes. 
In Fig. 12, we illustrate these effects:

 (i) The MDFs 
show a larger fraction of metal-rich ($-0.4$ $\le$ [m/H] $\le$ 0)
halo stars in the model with the larger bulge-to-disc-ratio (M2, $M_b =0.6$). 
This is because more of the metal-rich bulge stars can be
more efficiently  stripped into the stellar halo 
owing to its larger extension of the bulge in the model
with the larger bulge-to-disc-ratio. 
The MDF derived in the model with the smaller bulge-to-disc-ratio 
(M3, $M_b = 0$) seems to be more similar to the observed MDF of 
the NGC 5128 stellar halo, although we stress (see HH02) that the data
may be somewhat incomplete at the highest metallicities.
Nevertheless, the current comparisons would allow NGC 5128 to 
be formed by major merging between later-type spirals.

(ii) The stellar halo in the model (M4, GAS-type) which was generated
from progenitor spirals with the shallower disc metallicity gradients
(gaseous nebulae abundances; Zaritsky et al. 1994) 
can not match the observations.  Such models predict far too few
stars in the intermediate
metallicity range ($-1.4$ $\le$ [m/H] $\le$ $-0.6$).
The input discs with steeper metallicity gradients (OC-type) produce a
much closer match to the observations.
The opposite problem is presented by model M5, which has
a larger initial mean disc metallicity;
this model ends up predicting far too many halo stars
for [m/H] $\ge$ $-0.2$.
These results are the most prominent ones of our investigation, and
indicate that the detailed shape of the observed MDF in NGC 5128
can provide information on the mean metallicity and disc gradients in
the merger progenitor spirals (and thus, possibly, on their luminosity due to
the observed color-magnitude relation for disc galaxies).

(iii) The shapes of the MDFs of stellar haloes of elliptical galaxies
formed by major merging do not depend strongly on the orbital configurations 
of the merging (M6, 7, and 8). This probably reflects the fact that the physical
processes of violent relation and angular momentum redistribution
do not so strongly depend on the orbital parameters of major merging.
The PP model shows a slightly higher peak value of the MDF, because
a larger amount of disc stars can be stripped owing to
more efficient angular momentum redistribution process 
in this model.

(iv) Lastly, the stellar haloes in
the progenitor spirals exert noticeable effects on the 
stellar halo of the merger remnant 
at the low-metallicity end ([m/H] $\le$ $-1.2$).
The model M9 with the larger initial mass fraction of the spiral's halo (20 \%) 
produces too many halo stars at $-1.8$ $\le$ [m/H] $\le$ $-1$ 
and is not consistent with the observations.
The model haloes with mass fractions of 5\% yield much closer matches
to the data.  In short, the model comparisons appear to rule out 
progenitor spirals with very massive, metal-poor haloes.
%The model M10 with the lower peak value of the spiral's initial halo MDF
%([m/H] =  $-2$) also has difficulty in explaining the observed MDF 
%at  [m/H] $\sim$ $-1.6$ (M10).>>>?????<<< These results clearly suggest that 
%the detailed shape of the observed MDFs at metal-poor ranges
%([m/H] $\le$ $-1.4$) for the stellar halo of NGC 5128 
%can give some constraints on the MDFs of haloes of its progenitor spirals. 

\section{A new constraint on elliptical galaxy formation}

Although the observed tight color-magnitude relation
of elliptical galaxies 
has been considered to suggest
that elliptical galaxies  are old, coeval and homogeneous systems passively
evolving after the single initial burst of star formation associated 
with dissipative galaxy formation
(e.g., Bower, Lucey, \& Ellis 1992),
a growing number of recent observational results
suggest that elliptical galaxies exhibit a great variety of star formation history
(e.g., Worthey, Trager, \& Faber 1996).
The observed  narrowness of the fundamental plane of elliptical galaxies
implies a rather smaller range of permitted
%admitted 
dynamical state  
of the galaxies (Djorgovski \& Davis 1987; Dressler et al. 1987).
However, the morphological 
dichotomy between boxy-disky elliptical galaxies
(Kormendy \& Bender 1996) and departures of the projected density profiles
from the de Vaucouleurs $R^{1/4}$ law 
(Caon, Capaccioli, \& D'Onofrio 1993) 
show  a great variety of major stellar orbit families within the galaxies.
Elliptical galaxies thus show both
diversity and uniformity in their chemical, photometric and dynamical properties
which impose valuable constraints on any theoretical models of
elliptical galaxy formation.

The present study has demonstrated that, irrespective of bulge-to-disc-ratio
and MDF of of progenitor spirals,
elliptical galaxies formed by major merging show the tendency for 
stellar haloes to be populated predominantly by moderately metal-rich 
stars in the range $-1$ $<$ [m/H] $<$ 0 which were originally in the spiral
discs.  
%We propose that this generic trend in MDFs of elliptical galaxies formed by
%major merging can apply to elliptical galaxies other than NGC 5128.
Classical monolithic collapse scenarios
(e.g.,  Larson 1976; Carlberg 1984)
which have shown  strong negative metallicity gradients 
(i.e., more metal-poor in the outer part of a giant  elliptical) 
are less likely to yield the
moderately metal-rich populations in the outer halo regions that we see
in our test case of NGC 5128.
Therefore, if future systematic observations on MDFs of elliptical galaxies
confirm the above  generic trend in their halo MDFs,  it would provide
evidence more consistent with major merging as a promising scenario
of elliptical galaxy formation.

Spiral galaxies exhibit a color-magnitude (CM) relation
with a slope steeper than those of elliptical galaxies
(de Grijs \&  Peletier 1999). If the CM relation 
is due mostly to metallicity vs. luminosity  
(i.e., no significant age effects on integrated colors),
then more luminous spirals should be  more metal-rich.
The present numerical studies have demonstrated that 
the mean halo metallicity of an elliptical depends strongly
on the mean metallicity of the merger progenitor discs,
and furthermore that the mass fraction of stellar haloes
is not different between different merger models.
These observational and theoretical
results lead to the prediction that {\sl if} the majority of
E galaxies were formed by major mergers (that is, a single merger event between
roughly equal progenitors), there
should be a strong correlation between the color (mean metallicity)
of the stellar halo and the absolute magnitude of its host elliptical galaxy:
the more luminous (redder) elliptical should show redder halo colors.
In this picture, if there are very luminous ellipticals with 
rather metal-poor haloes that violate this correlation, they would
have been formed by a longer series of mergers from smaller, more metal-poor
progenitors or perhaps by accretion as in C\^ot\'e et al. (1998, 2002).
% WE DON'T UNDERSTAND THE SECOND POINT -- DIDN'T YOU ALREADY SAY THAT
% THE CM RELATION IS STEEPER FOR SPIRALS
%Secondly, there can be  CM relations of stellar haloes in elliptical galaxies with
%different luminosities: The brighter haloes show redder colors,
%and the slope of the CM relation should be more similar to that of spirals
%than to that of ellipticals. 

\subsection{Origin of the M31 stellar halo}

Since Mould \& Kristian (1986) first resolved clearly the 
brightest RGB and AGB stars in the M31 halo,
many observational studies of its MDF 
have been done (e.g., Pritchet \& van den Bergh 1988;
Christian \& Heasley 1991; Davidge 1993; Durrell, Harris, \& Pritchet 1994;
Couture et al. 1995;  Holland, Fahlman, \& Richer 1996;
Rich, Mighell \& Neil 1996; Reitzel, Guhathakurta, \& Gould 1998;
Reitzel \&  Guhathakurta 2001; Durrell, Harris, \& Pritchet 2001). 
One of the most remarkable results of these studies is that
the derived MDF of the M31 halo is very different from that of the Milky Way
halo.  All the studies listed above find that the M31 halo 
is dominated by moderately high-metallicity
([m/H] $\sim$ $-0.5$) populations 
at a very similar level to what we find in NGC 5128.
Durrell, Harris, \& Pritchet (2001) point out that the bulk of the
M31 halo stars are much less likely to come from dwarf galaxy populations,
just as for NGC 5128.
C\^ot\'e et al. (2000) demonstrated  that if the M31 halo was formed
by dissipationless merging of numerous subgalactic clumps with a 
luminosity ($L$) function of slope $dn/dL$ $\propto$ $L^{-1.8}$,
the observed MDF could be reproduced.
However, it is not clear {\sl a priori} why the MDFs of the Milky Way
and M31 should be so different in the same Local Group environment with
the plausible assumption that the surrounding pregalactic population of
clumps would be similar for both.
Ibata et al. (2001) suggest that the bulk of the M31 halo may have arisen
from a single event, namely cannibalization of stars from the 
similarly metal-rich satellite M32.
%Durrell, Harris,  \& Pritchet (2001) pointed out that
%the observed metal-rich stars with $Z$ = $0.3-0.5$ $Z_{\odot}$
%are much less likely to come from dwarf galaxy populations. 
%Thus it remains uncertain how the M31' halo acquired such metal-rich
%populations during its formation.

We here propose that the origin of the M31 stellar halo can be closely
associated with M31's bulge formation via major merging of two 
discs at an early epoch (see also Freeman 1999). 
In this scenario, the metal-rich halo stars were initially
within the outer parts of the merger progenitor spirals and tidally
stripped during the M31's bulge formation. This merger scenario
can give a plausible explanation as to why the MDFs in the haloes of
M31 and NGC 5128 resemble each other so strongly
(Harris \& Harris 2001).
If the observed difference in the metallicity of the dominant
metal-rich population between these two can reflect the difference
in mean metallicity of merger progenitor spirals between these two,
we can estimate the luminosity difference between the progenitor spiral
of M31's bulge and that of NGC 5128. 
The metallicity of the dominant halo population is [m/H] $\sim$ $-0.5$ 
for M31 (Durrell, Harris,  \& Pritchet 2001).
and $-0.4$ for NGC 5128 (HHP; HH00, 02). 
By using the observed luminosity-metallicity relation 
$Z$ $\propto$ $L^{0.4}$ (Mould 1984),
we can expect that  the possible luminosity difference of merger progenitor spirals
between these two big spheroidal components 
is  a factor of $\sim$ 2. 
This difference is appreciably smaller than 
the observed luminosity difference (a factor of $\sim$ 4) between M31's bulge 
with $L_{V}$ $\sim$ 1.1 $\times$  $10^{10}$ $L_{\odot}$ corresponding to
about 40 \% of total stellar light of M31 (van den Bergh 2000) 
and  NGC 5128
with $L_{V}$ $\sim$ 4.7 $\times$  $10^{10}$ $L_{\odot}$ (Hui et al. 1995).

%  THE PARAGRAPH ABOUT GLOBULAR CLUSTERS REPEATED WHAT WAS SAID EARLIER

Freeman (1999) suggested that the big M31 bulge could be formed
by a past merger event on the basis of M31's halo structural
properties. He also pointed out that if M31's bulge was formed by merging,
its {\it outer} stellar halo components should show significant rotation
owing to efficient angular momentum transfer during merging. Our present
simulations have confirmed that the outer stellar halo components
clearly show rotation after major merging.
We note that Perrett et al. (2002) have indeed found a strong rotation signature for
both types of globular clusters in the M31 halo (160 km s$^{-1}$ for the
metal-rich population, 130 km s$^{-1}$ for the metal-poor population),
extending outward to $\sim 15-20$ kpc.

Big spheroidal bulges (like M31's) are suggested to be physically  
different from  box- and peanut-shaped bulges in spirals
(e.g., Wyse et al 1997 and references therein): Spheroidal bulges are ``true'' bulges
similar to giant ellipticals
whereas the  box- and peanut-shaped bulges are edge-on bars formed through dynamical
instability of discs (e.g., Combes \& Sanders 1981).
We suggest that if ``true''  spheroidal bulges are formed by merging
(rather than disc instability),
spirals with such big bulges are more likely to show halo MDFs similar to
that of M31 rather than to that of the Milky Way.
It would be an interesting future study to correlate the MDFs of stellar haloes
with bulge shapes in different Hubble types of galaxies. 

\subsection{Discussion and Remaining Problems}

We close by briefly mentioning alternate models of halo formation applicable
to NGC 5128 and other ellipticals, along with difficulties still faced by most
interpretations.

E galaxies may also form by longer series of  mergers (accretion)
of smaller subgalactic clumps.
The dissipationless accretion of dwarf galaxies
with appropriate chosen luminosity functions can be also responsible
for MDFs with relatively metal-rich peaks (e.g., C\^ot\'e et al 1998, 2000, 2002).
As also noted earlier, Beasley et al. (2002a,b) take the alternate route 
of hierarchical merging of pregalactic gas clouds to model the NGC 5128 MDF, with
first-order matches possible in a wide variety of circumstances.
Without repeating the extensive analysis in these papers, we note only that
their work demonstrates that a unique physical interpretation for the 
observed MDFs may  still not be in hand.  Eventually, all the observational
parameters including kinematics and halo/bulge structures 
as well as MDFs must be brought to bear on the question.

There are, however, two interesting interpretive problems which 
the globular cluster populations confront
us with.  First is the sheer size of the cluster population in NGC 5128.
Coincidentally, the luminosity of NGC 5128 ($M_V \simeq -22.0$)
is similar to what we would
get if we simply combined two galaxies like M31, or M31 plus the Milky Way
(Pritchet \& van den Bergh 1999).  Thus a simple merger of the two
would give an elliptical of about the same size as NGC 5128 and
with $\sim 600 - 800$ clusters.   A GC population of that size is two to three times
smaller than the $\sim 1500 - 2000$ globular clusters that are in
NGC 5128 (G.Harris et al. 1984).  That is, NGC 5128 has a GC specific frequency
significantly higher than normal spiral galaxies. Contrary to many 
comments in the literature, this classic ``specific frequency problem''
cannot be solved by invoking the presence of gas and then
forming large numbers of clusters during the merger.
This is because field stars will also form from the gas, and the ratio of
field stars to GCs will stay at much the same level.  See the discussion
of Harris (2001), who notes that to produce noticeable changes in the 
specific frequency, it is necessary to have {\sl both} extremely large
amounts of gas {\sl and} extremely high cluster formation efficiency.
Under such conditions, we would no longer be talking about a simple disc/disc
merger but rather an earlier epoch of primary galaxy formation from
hierarchical merging of gas clouds.

A second concern relates to the spatial distributions of the GCs in the metal-poor
and metal-rich subgroups that many or most giant ellipticals contain.
The GCs in the metal-rich mode are not found only in the bulge region as noted
above, but also extend outward far into the halo in significant numbers
(e.g., G.Harris et al. 1992; Lee \& Geisler 1993; Geisler, Lee, \& Kim 1997).
Such metal-rich clusters are also present in the outer halo of M31 
(Perrett et al. 2002).
Within the context of the merger scenario we have tested here, it is not
clear how these metal-rich, outer-halo clusters arise.  Clusters formed
from appropriately metal-rich gas during the merger are likely to build
up in the central regions and add to the bulge population.  
Metal-rich globular clusters that are {\sl already} 
distributed widely throughout the discs or haloes 
of the progenitor spirals would end up populating the halo of the
emergent elliptical.  Such clusters are present to some extent in M31,
but not in the Milky Way.  In summary, it is still difficult for any
one model to correctly predict the total ensemble of features of
the giant ellipticals, which may have formed by a combination of routes.

\section{Conclusions}

We have numerically investigated MDFs of stellar haloes of
elliptical galaxies formed by major merging in order
to elucidate the origin of the observed MDF of NGC 5128. We mainly investigated
the MDF of halo stars with $R$ $>R_{\rm d}$ (where $R$ and $R_{\rm d}$ are the radius
from the center of a merger remnant and the initial size of the merger progenitor spiral)
in the merger remnant for several models with different physical parameters
of the progenitor spirals. Our numerical study clearly demonstrates the importance
of {\it dynamical processes of galaxy merging} (i.e., tidal stripping of
metal-poor disc stars) in determining MDFs of stellar haloes of elliptical galaxies. 
We summarize our principal results as follows.

(1) The stellar halo of an elliptical galaxy formed by major merging
is predominantly populated by moderately metal-rich stars 
with [m/H] $\sim$ $-0.4$, which come from 
the outer parts of the discs of the merging spirals.
Furthermore, the fraction of metal-poor stars with [m/H] $<$ $-1.0$
in the halo is very small ($\sim$ 17 \%).
The MDF of the  merger remnant does not show
a strong radial dependence for 0.5 $<$ $R$ $<$ 2.0 (in our units). 
These results are broadly consistent with
recent {\it HST} observations on MDFs of the stellar halo of NGC 5128.

(2) Irrespective of the physical parameters of merger progenitor discs,
the MDFs of stellar haloes of merger remnants
show both the dominant moderately metal-rich populations
and the minor metal-poor ones. 
Thus future systematic  observations on MDFs of elliptical galaxies
can assess the relative importance of major galaxy merging in
the formation of elliptical galaxies by confirming  
the above generic trend of MDFs.

(3) The MDFs of elliptical galaxies formed by major merging
depend  strongly on the initial MDFs of merger progenitor discs,
because halo stars are dominated by those which are initially in the outer part
of the discs and tidally stripped during merging.

(4) MDFs of stellar haloes do not depend 
so strongly on the MDFs of the progenitor spiral bulges,
essentially because bulge stars are not be tidally
stripped so efficiently during merging owing to bulge's compact
configuration. However, the shapes of MDFs of a merger remnant
at $-0.2$ $\le$ [m/H] $\le$ 0 can be affected by the bulge-to-disc-ratio  
of the merger progenitor spiral.

(5) The shape of the halo MDF of a merger remnant at [m/H] $\le$ $-1.0$ 
depends on the mass fraction and the mean metallicity of the progenitor spiral's halo.
Therefore the observed shape of the halo MDF at [m/H] $\le$ $-1.0$ in NGC 5128 
can give some constraints on the stellar populations of the progenitor 
spiral's stellar halo. 

(6) The observed similarity in MDFs of stellar haloes between M31 and NGC 5128
can be naturally explained, if the M31 bulge was also formed by major merging
between two spirals with the masses (a factor of $2-4$)  smaller than     
those that are assumed to merge to from NGC 5128 in the present study.
The observed markable difference in MDFs between stellar haloes and GCs 
for NGC 5128 and M31 can be also explained by the merger scenario.

(7) Our simulations predict that if most elliptical galaxies are formed
by major merging, the photometric properties (e.g., colors and magnitudes)
of a stellar halo in an elliptical galaxy
can correlate with those of the galaxy (i.e., redder Es have redder stellar haloes).
Also our study predicts that there can be CM relations in stellar haloes of elliptical
galaxies. 

%\acknowledgment
\section{Acknowledgment}
K. B.  acknowledges the financial support of the Australian Research Council
throughout the course of this work.  W.E.H. and G.L.H.H. acknowledge
financial support from the Natural Sciences and Engineering Research Council
(NSERC) of Canada and the hospilatity and support at Mt Stromlo Observatory
(RSAA/ANU) during research leaves when this paper was written.

\begin{figure}
%\vspace{11pc}
%\psfig{file=f1.ps}
%\psfig{file=f1.ps}
%\psfig{file=f1.eps,width=18.5cm}
\psfig{file=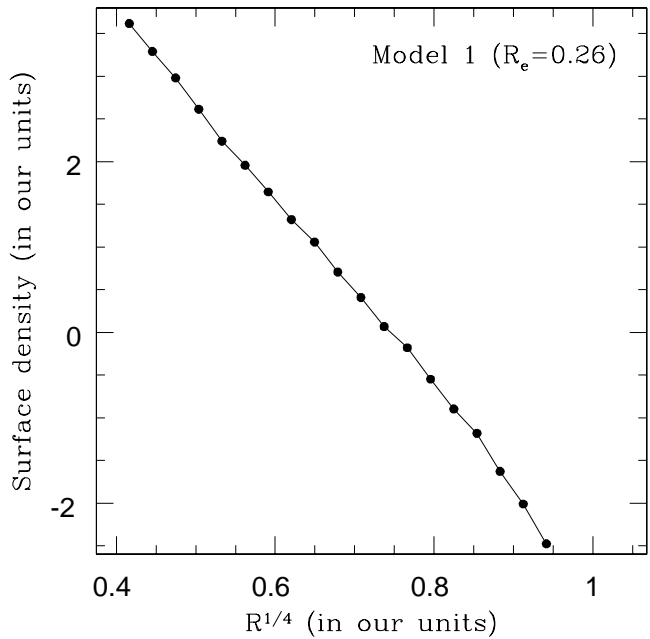,width=8.5cm}
\caption{
Surface density profile of the model M1. The profile
of the simulated elliptical galaxy is well described as
the $R^{1/4}$ law (de Vaucouleurs law) with the effective
radius ($R_{\rm e}$) of 0.26 (${R_{\rm e}}^{1/4}$ = 0.71)
in our units (corresponding to 4.6 kpc) in this model.
This simulated profile is consistent with
the observed one by van den Bergh (1976) for NGC 5128. 
}
\label{Figure. 13}
\end{figure}
\appendix
\section[]{Light profiles of merger remnants}

Although the main purpose of this paper is not to discuss
structural properties of NGC 5128, it is important
for the present study to confirm  whether our numerical  simulations
can reproduce not only the MDF observed in the NGC 5128  stellar halo  
but also its global structural properties. 
 In the following, we assume that 
(1) the distance ($d$) of NGC 5128 is 3.9 Mpc (which is the same
as that adopted in HH00 and HH02) 
and (2) the simulated  density profile corresponds to the light profile
(i.e., mass-to-light-ratio is constant for the entire region of a simulated
elliptical galaxy).
We then compare the  observed radial brightness profile  of NGC 5128 
by van den Bergh (1976) with the simulated ones in models with
initial dynamical conditions different with one another
(i.e., M1, 2, 3, 7, and 8).

van den Bergh (1976) found that 
(1) the effective radius of NGC 5128 is about $5\farcm 5$
or $\sim$ 6.2 kpc for $d$ = 3.9 Mpc  
and (2) the radial brightness distribution follows de Vaucouleurs's law.
The effective radius of the simulated elliptical galaxy
in each model is 4.6 (M1), 4.6 (M2), 5.3 (M3), 6.1 (M7), and 7.0 kpc (M8).
These derived values are not significantly different from the observed one.
 As is shown in Fig. 13,
the simulated radial surface density profile in the model M1 follows  
de Vaucouleurs's law. The radial surface density profiles   
in other models can be also described as de Vaucouleurs's ones,
which is consistent with
the above observation (2). These comparison thus suggest
that our model can reproduce the observed structural properties
of NGC 5128 reasonably well.

%%%%%%TABLE

\begin{table*}
\centering
\begin{minipage}{185mm}
\caption{Model parameters}
\begin{tabular}{ccccccc}
Model no.  
& $M_{\rm b}$  
& metallicity gradient 
& ${\rm [m/H]}_{\rm d}$  
& orbit  
& $f_{h}$ 
& ${\rm [m/H]}_{\rm h}$ \\
M1 & 0.2 & OC & -0.15 & PR & 0.05 & -1.4 \\
M2 & 0.6 & OC & -0.15 & PR & 0.05 & -1.4 \\
M3 & 0.0 & OC & -0.15 & PR & 0.05 & -1.4 \\
M4 & 0.2 & GAS & -0.15 & PR & 0.05 & -1.4 \\
M5 & 0.2 & OC & 0.0 & PR & 0.05 & -1.4 \\
M6 & 0.0 & OC & -0.15 & PR & 0.05 & -1.4 \\
M7 & 0.0 & OC & -0.15 & PP & 0.05 & -1.4 \\
M8 & 0.0 & OC & -0.15 & RR & 0.05 & -1.4 \\
M9 & 0.2 & OC & -0.15 & PR & 0.20 & -1.4 \\
M10 & 0.2 & OC & -0.15 & PR & 0.05 & -2.0 \\

\end{tabular}
\end{minipage}
\end{table*}

%%%%%%FIGURE CAPTION

\newpage

%\newpage
%\begin{figure}
%\vspace{11pc}
%\caption{$P(>x_{\rmn{gap}})$ as a function of $x_{\rmn{gap}}$ for, 
% from left to right, $N=160$, 150, 140, 110, 100, 90, 50, 45 and~40.}
%\label{appenfig}
%\end{figure}

%%%%%%%%%%%%%%%%%%%%%%%%%%%%%%%%%%%%%%%%%%%%%%

%\newpage

%\begin{figure*}
%%\vspace{11pc}
%\caption{ 
%}
%\label{Figure. }
%\end{figure*}


\begin{thebibliography}{99}


\bibitem{}Beasley, M.A., Baugh, M.A., Forbes, D.A., Sharples, R.M.,
and Frenk, C.S. 2002a, MNRAS, 333, 383 

\bibitem{} Beasley, M.A., Harris, W.E., Harris, G.L.H., and Forbes, D.A.
2002b, MNRAS, accepted (astro-ph/0211586) 

\bibitem{} Bekki, K., Beasley, M.A., Forbes, D.A., \& Couch, W. J.,
2002, accepted in MNRAS

\bibitem{} Bower, R. G., Lucey, J. R., \& Ellis, R. S. 1992, MNRAS, 254, 601 

\bibitem{} Brodie, J. P., \&  Huchra, J. P. 1991, 379, 157


\bibitem{}
Caon, N., Capaccioli, M., \& D'onofrio, M. 1993, MNRAS, 265, 1013

\bibitem{}
Carlberg, R. G. 1984, ApJ, 286, 416

\bibitem{} Chiba,~M., \& Yoshii,~Y. 1998, AJ, 115, 168 

\bibitem{} Christian, C. A., \&  Heasley, J. N. 1986, ApJ, 303, 216 


\bibitem{} Combes, F.; Sanders, R. H. 1981, A\&A, 96, 164


\bibitem[]{}  C\^ote,~P., Marzke,~R.~O., West,~M.~J., 1998, ApJ, 501, 554 

\bibitem[]{}  C\^ote,~P., Marzke,~R.~O., West,~M.~J., \& Minniti,~D.
2000, ApJ, 533, 869

\bibitem[]{}  C\^ote,~P., West,~M.~J., Marzke,~R.~O. 2002, ApJ, 567,853

\bibitem{} Couture, J, Racine, R.,  Harris, W. E., \&  Holland, S. 
1995, AJ, 109, 2050

\bibitem{} Davidge, T. J. 1993, ApJ, 409, 190

\bibitem{dgp99}
de Grijs, R., \& Peletier, R. F., 1999, MNRAS, 310, 157

\bibitem{}
Djorgovski, S., \& Davis, M. 1987, ApJ, 313, 59

\bibitem{}
Dressler, A., Lynden-Bell, D., Burstein, D., Davies, R. L., Faber, S. M.,
Terlevich, R. J., \& Wegner, G. 1987, ApJ, 313, 42

\bibitem{} Durrell, P. R., Harris, W. E., \& 
 Pritchet, C. J. 1994, AJ, 108, 2114

\bibitem{} Durrell, P. R., Harris, W. E., \& 
 Pritchet, C. J. 2000, AJ, 121, 2557


\bibitem{} Elson, R. A. W. 1997, MNRAS, 286, 771

\bibitem{fe80} Fall, S. M., \& Efstathiou, G., 1980, MNRAS, 193, 189 

\bibitem{fo97} Forbes, D. A., Brodie, J. P., Grillmair, C. J., 1997, AJ, 113, 1652  

\bibitem{fr70} Freeman, K. C., 1970, ApJ, 160, 811

\bibitem[]{} Freeman,~K.~C. 1987, ARA\&A, 25, 603

\bibitem[]{} Freeman,~K.~C. 1999, in  The Third Stromlo Symposium: 
The Galactic Halo, eds. Gibson, B.K., Axelrod, T.S. \& Putman, M.E., 
ASP Conference Series Vol. 165, p. 167

\bibitem{} Friel, E. D. 1995, ARAA, 33, 381

\bibitem{} Frogel, J. A., Tiede, G. P., \&  Kuchinski, L. E. 1999, AJ, 117, 2296

\bibitem{} Geisler, D., Lee, M. G., \& Kim, E. 1996, 111, 1529

\bibitem{} Grillmair, C. J., Lauer, T. R., Worthey, G., Faber, S. M.,
Freedman, W. L., Madore, B. F., Ajhar, E. A., Baum, W. A.,
Holtzman, J. A., Lynds, C. R., O'Neil, E. J. Jr., 
\& Stetson, P. B. 1996, AJ, 112, 1975

\bibitem{} Harris, G. L. H., Hesser, J. E., Harris, H. C., \& Curry, P. J.
1984, ApJ, 287, 175

\bibitem{} Harris, G. L. H., Geisler, D., Harris, H. C.,
\& Hesser, J. E. 1992, AJ, 104, 613

%\bibitem{ha91}  Harris, W. E., \& van den Bergh, S., 1981, AJ, 86, 1627 

\bibitem{ha91} Harris, W. E., 1991, ARAA, 29, 543 

\bibitem{} Harris, W. E., 2001, in Star clusters,
 Saas-Fee Advanced Course 28. Lecture Notes 1998, Swiss Society 
for Astrophysics and Astronomy. 
Edited by L Labhardt and B. Binggeli. Published by Springer-Verlag, Berlin, 2001,  p223.

\bibitem{} Harris, W. E., Harris, E. H., \& Poole, G.  B. 1999, AJ, 117, 855 (HHP) 

\bibitem{} Harris, G. L. H., \& Harris, W. E., 2000, AJ, 120, 2423 (HH00)

\bibitem{} Harris, W. E., \& Harris, G. L. H., 2001, AJ, 122, 3065

\bibitem{} Harris, W. E., \& Harris, G. L. H., 2002, AJ, 123, 3108 (HH02)


\bibitem{} Han, M.; Hoessel, J. G., Gallagher, J. S., III,  Holtsman, J.,
Stetson, P. B. 1997, AJ, 113, 1001 

\bibitem{} Holland, S., Fahlman, G. G., \&  Richer, H. B. 
1996, 112, 1035

\bibitem{} Hui, X., Ford, H. C., Freeman, K. C., 
Dopita, M.  A. 1995, ApJ, 449, 592

\bibitem{} Ibata, R.,  Irwin, M., Lewis, G., Ferguson, A. M. 
N., \&  Tanvir, N. 2001, Nature, 412, 49

\bibitem{}
Kormendy, J., \& Bender, R. 1996, ApJL, 464, 119


\bibitem{} Larsen, S. S., Brodie, J. P., Huchra, J. P.,
Forbes, D. A., Grillmair, C. J. 2001, AJ, 121, 2974 

\bibitem{} Larson, R. B. 1975, MNRAS, 173, 671

\bibitem{}  Lee, M. G., \& Geisler, D. 1993, AJ, 106, 493


\bibitem[]{} Majewski,~S.~R. 1993, ARA\&A, 31, 575

\bibitem[]{} Marleau, F. R., Graham, J. R., Liu, M. C., \&
Charlot, S. 2000, AJ, 120, 1779

\bibitem[]{} Martinez-Delgado, D., \&  Aparicio, A. 1998, AJ, 115, 1462

\bibitem[]{} McWilliam, A., \&  Rich, R. M. 1994, ApJS, 91, 749

\bibitem{} Minniti, D. 1996, ApJ, 459, 175 

\bibitem{} Mould, J. 1984, PASP, 96, 773

\bibitem{} Mould, J., \&  Kristian, J. 1986, ApJ, 305, 591 

\bibitem{} Peng, E. W., Ford, H. C., \&  Freeman, K. C.  2002, in preparation


\bibitem{} Perrett, K. M., Bridges, T. J., Hanes, D. A., Irwin, M. J.,
Brodie, J. P., Carter, D., Huchra, J. P., \&  Watson, F. G.
2002, AJ, 123, 2490 

\bibitem{} Pritchet, C. J., \&  van den Bergh, S. 1999, AJ, 118, 883


\bibitem{} Reitzel, D. B., Guhathakurta, P., \&  Gould, A. 1998, AJ, 116, 707

\bibitem{} Reitzel, D. B.,  \& Guhathakurta, P. 2000,
The Galactic Halo : From Globular Cluster to Field Stars, 
Proceedings of the 35th Liege International Astrophysics Colloquium, held 5-8 July, 1999. 
Edited by A. Noels, P. Magain, D. Caro, E. Jehin, G.
Parmentier, and A. A. Thoul. (Liege, Belgium : 
Institut d'Astrophysique et de Geophysique, 2000.),  p.365


\bibitem{}  Rejkuba, M., Minniti, D., Courbin, F., \&  Silva, D. R. 
2002, ApJ, 564, 688

\bibitem{} Rich, R. M.,  Mighell, K. J., \&  Neill, J. D.
Formation of the Galactic Halo...Inside and Out, ASP Conference Series, Vol. 92, 
1996, Heather Morrison and Ata Sarajedini, eds., p. 544. 

\bibitem{} Searle,~L. \& Zinn,~R. 1978, ApJ, 225, 357 

\bibitem{} Soria, R. et al. 1996, ApJ, 465, 79 

\bibitem{} Sugimoto,~D., Chikada,~Y., Makino,~J., Ito,~T., Ebisuzaki,~T., \&
Umemura, M. 1990,  Nature, 345, 33

%\bibitem{to64}
%Toomre, A., 1964, ApJ, 139, 1217

\bibitem{too77}
Toomre, A., 1977, in  The evolution of galaxies and stellar Populations
ed. by B. Tinsley \& R. Larson (New. Haven. CN: Yale Univ. Press), p401

\bibitem[]{} van~den~Bergh,~S. 1976, AJ,  208, 673 

\bibitem[]{} van~den~Bergh,~S. 1996, PASP,  108, 986

\bibitem{} van den Bergh, S., 1999, preprint (astro-ph/9908050)

\bibitem{} van den Bergh, S., 2000, in the galaxies in the Local group. 


\bibitem{wi77}
Wielen, R., 1977, A\&A, 60, 263


\bibitem[]{}
Worthey, G., Trager, S. C., \& Faber, S. M. 1996,
ed. A.Buzzoni, A.Renzini, and A.Serrano, ASP Conf.Ser. Vol.86, p. 203


\bibitem{} 
 Wyse, R. F. G., Gilmore, G., \&  Franx, M. 1997, ARAA, 35, 637

\bibitem{za94}
Zaritsky, D., Kennicutt, R. C., Huchra, J. P. 1994, ApJ, 420, 87


\end{thebibliography}
\end{document}